\documentclass[12pt,letterpaper]{article}
\usepackage{amsmath,amsfonts,amsthm,amssymb,stmaryrd,bm,cite} 
\usepackage{relsize,tabls,mathtools}
\usepackage{array}   

\allowdisplaybreaks  

\theoremstyle{plain}
\numberwithin{equation}{section}

\newcommand{\integers}{{\mathbb Z}}
\newcommand{\complex}{{\mathbb C}}
\newcommand{\hscript}{{\mathcal H}}
\newcommand{\sscript}{{\mathcal S}}
\newcommand{\sscripthat}{\widehat{\sscript}}
\newcommand{\alphabar}{\overline{\alpha}}
\newcommand{\betabar}{\overline{\beta}}
\newcommand{\deltabar}{\overline{\delta}}
\newcommand{\gammabar}{\overline{\gamma}}
\newcommand{\jbar}{\overline{J}}

\newcommand{\bfp}{\mathbf{p}}       
\newcommand{\bfx}{\mathbf{x}}       

\newcommand{\ab}[1]{\left|#1\right|}
\newcommand{\doubleab}[1]{\left|\left|#1\right|\right|}
\newcommand{\brac}[1]{\left\{#1\right\}}
\newcommand{\paren}[1]{\left(#1\right)}
\newcommand{\sqbrac}[1]{\left[#1\right]}

\newcommand{\ket}[1]{{\left|#1\right>}}
\newcommand{\bra}[1]{{\left<#1\right|}}

\begin{document}
\title{TOY MODELS FOR\\QUANTUM FIELD THEORY
}
\author{S. Gudder\\ Department of Mathematics\\
University of Denver\\ Denver, Colorado 80208, U.S.A.\\
sgudder@du.edu
}
\date{}
\maketitle

\begin{abstract}
In order to better understand quantum field theory we present some toy models on finite dimensional Hilbert spaces. We discuss how these models converge to a discrete spacetime version of quantum field theory. We first define toy fermion, boson and mixed fermion-boson free quantum fields. The free fields are employed to form what we consider to be the simplest examples of interacting fields which we then use to construct Hamiltonian densities. Scattering operators are defined in terms of these densities and are employed to construct scattering probabilities. Although our definition of a scattering operator is not the usual ``second-quantization'' definition we believe it is simpler and more natural. Emphasis is placed on finding the eigenstructures of toy quantum fields.
\end{abstract}\newpage

\section{Introduction}  
Many experts admit that quantum field theory is a difficult and complicated subject. Besides that, it is plagued with infinities, divergences and mathematical inconsistencies. The purpose of this work is to try to gain a deeper understanding of the foundations and basic principles of the theory by considering some toy models. By necessity, these toy models will involve over-simplifications and they do not directly apply to physical reality. However, these models exist on finite-dimensional Hilbert spaces and are mathematically rigorous. Moreover, it is possible that, in some sense, they converge to an approximation of the real observed world. In Section~5 we discuss ways this may be accomplished. In particular, we assume that special relativity is an approximation to a discrete spacetime which is the fundamental description of the structure of the universe \cite{bdp16,cro16,gud172,hag14,hoo14}.

We first define free fermion, boson and mixed fermion-boson quantum fields. Applying energy and particle number cut-offs, the free fields become rigorous operators on finite-dimensional Hilbert spaces. These operators are the bases for our toy models. The free fields are employed to form what we consider to be the simplest examples of interacting fields. Many examples for free and interacting toy quantum fields are given in Sections~3 and~4. To better understand these fields we stress the eigenstructure of their quantum operators. In Section~5 we use interacting fields to construct Hamiltonian densities. Scattering operators, which are the most important operators of quantum field theory, are defined in terms of the Hamiltonian densities. We finally show how scattering probabilities can be found using the scattering operators. Scattering cross-sections, decay probabilities and particle lifetimes can also be computed in the standard ways.

It is hoped that these computed quantities will converge to numbers that will agree with experiment, but this will have to wait for future work. As we shall show in examples, even these toy models quickly involve complications which can only be solved with computer assistance. We warn the reader that our definition of a scattering operator is not the standard one involving ``second-quantization''. Perhaps it is too na\'ive, but we believe that this method is simpler and more natural.

\section{Free Toy Quantum Fields} 
In this toy model we consider a system with at most $s$ particles $p_1,p_2,\ldots ,p_s$. We first assume that the particles are all fermions with the same mass (say, electrons). For simplicity, we neglect spin since the model can be extended to include spin in a straightforward way. We construct a finite-dimensional complex Hilbert space $K^s$ with orthonormal basis
\begin{equation*}
\ket{0},\ket{p_1},\ldots ,\ket{p_s}\cdots,\ket{p_{i_1}\ldots p_{i_n}},\cdots ,\ket{p_1p_2\ldots p_s}
\end{equation*}
where $\ket{p_{i_1}\ldots p_{i_n}}$ represents the state in which there are $n$ particles with energy-momentum $p_{i_1},\ldots ,p_{i_n}$. The
\textit{vacuum state} with no particles is $\ket{0}$. This should not be confused with the zero vector in $K^s$ denoted by $0$. The basis is antisymmetric in the sense that if two particles are interchanged, then a  negative sign results. For example, we have that
$\ket{p_1p_2p_3}=-\ket{p_3p_2p_1}$. It follows that no two entries in a basis vector agree. Notice that the dimension of $K^s$ is
\begin{equation*}
\dim K^s=\sum _{j=0}^s\binom{s}{j}=2^s
\end{equation*}
For a particle $p_j$ we define the \textit{annihilation operator} $a(p_j)$ on a basis element by
\begin{equation*}
a(p_j)\ket{p_jp_{i_1}\dots p_{i_n}}=\ket{p_{i_1}\ldots p_{i_n}}
\end{equation*}
and if $\ket{p_{i_1}\ldots p_{i_n}}$ does not contain an entry $p_j$, then
\begin{equation*}
a(p_j)\ket{p_{i_1}\ldots p_{i_n}}=0
\end{equation*}
We then extend $a(p_j)$ to $K^s$ by linearity. Notice for example, that $a(p_1)\ket{p_1p_2}=\ket{p_2}$ while
\begin{equation*}
a(p_1)\ket{p_2p_1}=-a(p_1)\ket{p_1p_2}=-\ket{p_2}
\end{equation*}
We interpret $a(p_j)$ as the operator that annihilates a particle with energy-momentum $p_j$.

The adjoint $a(p_j)^*$ of $a(p_j)$ is the operator on $K^s$ defined by
\begin{equation*}
a(p_j)^*\ket{p_jp_{i_1}\ldots p_{i_n}}=0
\end{equation*}
and if $\ket{p_{i_1}\ldots p_{i_n}}$ does not contain an entry $p_j$, then
\begin{equation*}
a(p_j)^*\ket{p_{i_1}\ldots p_{i_n}}=\ket{p_jp_{i_1}\ldots p_{i_n}}
\end{equation*}
We call $a(p_j)^*$ a \textit{creation operator} and interpret $a(p_j)^*$ as the operator that creates a particle with energy-momentum $p_j$. Defining the \textit{commutator} and \textit{anticommutator} of two operators $A,B$ on $K^s$ by $[A,B]=AB-BA$ and $\brac{A,B}=AB+BA$, respectively, it is easy to check that $a(p_j)$ and $a(p_j)^*$ have the characteristic properties:
\begin{align}         
\label{eq21}
\brac{a(p_i),a(p_j)}&=\brac{a(p_i)^*,a(p_j)^*}=0\\
\label{eq22}         
\brac{a(p_i),a(p_j)^*}&=\delta _{ij}I
\end{align}
where $I$ is the identity operator. For example, if $i\ne j$ then
\begin{align*}
\brac{a(p_i),a(p_i)^*}\ket{p_j}&=a(p_i)a(p_i)^*\ket{p_j}+a(p_i)^*a(p_i)\ket{p_j}\\
   &=a(p_i)\ket{p_ip_j}+a(p_i)^*(0)=\ket{p_j}
\end{align*}
while if $i=j$, then
\begin{align*}
\brac{a(p_i),a(p_i)^*}\ket{p_j}&=a(p_i)a(p_i)^*\ket{p_j}+a(p_i)^*a(p_i)\ket{p_j}\\
   &=a(p_i)(0)+a(p_i)^*\ket{0}=\ket{p_i}
\end{align*}
Of course, it follows that $a(p_i)$ and $a(p_j)$ do not commute when $i\ne j$.

Corresponding to a fermion $p_j$ and a complex number $\alpha _j\in\complex$ we define the \textit{annihilation-creation operator}
(AC-\textit{operator}) $\eta (p_j)$ by
\begin{equation*}
\eta (p_j)=\alpha _ja(p_j)+\alphabar _ja(p_j)^*
\end{equation*}
It is clear that $\eta (p_j)$ is a self-adjoint operator on $K^s$ and it follows from \eqref{eq21} and \eqref{eq22} that
\begin{equation*}
\brac{\eta (p_i),\eta (p_j)}=2\ab{\alpha _i}^2\delta _{ij}I
\end{equation*}
It follows that $\eta (p_i)$ and $\eta (p_j)$ do not commute when $i\ne j$. If $p_1,\ldots ,p_n$ are distinct fermions in our system and
$\alpha _j\in\complex$, $j=1,\ldots ,n$ we call
\begin{equation}         
\label{eq23}
\phi =\sum _{j=1}^n\eta (p_j)=\sum _{j=1}^n\sqbrac{\alpha _ja(p_j)+\alphabar _ja(p_j)^*}
\end{equation}
a \textit{free fermion toy quantum field}. Again, $\phi$ is a self-adjoint operator on $K^s$.

We now consider a system of bosons with energy-momentum $q_j$, $j=1,\ldots ,n$ and having the same mass (say, photons). We again have the restriction that there are at most $s$ particles. In this case, we can have that $q_i=q_j$ so some of the particles can be identical. We construct the symmetric Hilbert space $J^{(n,s)}$ with orthonormal basis $\ket{q_{i_1}\cdots q_{i_k}}$. We sometimes use the notation
\begin{equation*}
\ket{q_{i_1}^{j_1}q_{i_2}^{j_2}\cdots q_{i_k}^{j_k}}
\end{equation*}
to represent the state in which there are $j_1$ bosons of type $i_1,\ldots ,j_k$ bosons of type $i_k$. For example
\begin{equation*}
\ket{q_1^2q_2^0q_3^3q_4}=\ket{q_1q_1q_3q_3q_3q_4}
\end{equation*}
Since $J^{(n,s)}$ is symmetric, an interchange of entries does not affect the state. For example,
\begin{equation*}
\ket{q_1q_2q_3}=\ket{q_2q_1q_3}=\ket{q_2q_3q_1}
\end{equation*}
Notice that there is a one-to-one correspondence between basis elements and multisets with at most cardinality $s$ that have at most $n$ different elements. It is well-known that the number of multisets with $k$ elements chosen from among $q_1,\ldots ,q_n$ is 
\begin{equation*}
{n+k-1\choose k}=\frac{(n+k-1)!}{k!(n-1)!}
\end{equation*}
We conclude that
\begin{equation*}
\dim J^{(n,s)}=\sum _{k=0}^s{n+k-1\choose k}
\end{equation*}
For example, if $n=2$, $s=3$ we have that
\begin{equation*}
\dim J^{(2,3)}=\sum _{k=0}^3{k+1\choose k}={1\choose 0}+{2\choose 1}+{3\choose 2}+{4\choose 3}=1+2+3+4=10
\end{equation*}
The basis elements for $J^{(2,3)}$ are
\begin{equation*}
\ket{0},\ \ket{q_1},\ \ket{q_2},\ \ket{q_1^2},\ \ket{q_1q_2},\ \ket{q_2^2},\ \ket{q_1^3},\ \ket{q_1^2q_2},\ \ket{q_1q_2^2},\ \ket{q_2^3}
\end{equation*}

For a boson $q_j$ we define the \textit{annihilation operator} $a(q_j)$ by
\begin{equation*}
a(q_j)\ket{q_j^kq_{j_1}^{k_1}\cdots q_{j_t}^{k_t}}=\sqrt{k\,}\,\ket{q_j^{k-1}q_{j_1}^{k_1}\cdots q_{j_t}^{k_t}} 
\end{equation*}
where $k+k_1+\cdots +k_t\le s$. The corresponding \textit{creation operator} $a(q_j)^*$ is given by
\begin{equation*}
a(q_j)^*\ket{q_j^kq_{j_1}^{k_1}\cdots q_{j_t}^{k_t}}
  =\begin{cases}\sqrt{k+1}\,\ket{q_j^{k+1}q_{j_1}^{k_1}\cdots q_{j_t}^{k_t}}&\hbox{if }k+k_1+\cdots +k_t<s\\
  0&\hbox{if }k+k_1+\cdots +k_t=s\end{cases}
\end{equation*}
As before, these operators are extended to $J^{(n,s)}$ by linearity. We define the boundary $\jbar ^{(n,s)}$ of $J^{(n,s)}$ as the subspace of $J^{(n,s)}$ generated by the basis vectors
\begin{equation*}
V^{(n,s)}=\brac{\ket{q_{i_1}^{j_1}\cdots q_{i_t}^{j_t}}\colon j_1+\cdots +j_t=s}
\end{equation*}
We have that
\begin{align*}
\sqbrac{a(q_j),a(q_k)}&=\sqbrac{a(q_j)^*,a(q_k)}=0\ \hbox{on } J^{(n,s)}\\
\sqbrac{a(q_j),a(q_k)^*}&=\delta _{jk}I\ \hbox{   on }J^{(n,s)}\smallsetminus\jbar ^{(n,s)}
\end{align*}
while on $\jbar ^{(n,s)}$ we have that
\begin{equation*}
\sqbrac{a(q_1),a(q_k)^*}\ket{\psi}
  =\begin{cases}-\sqrt{N_{q_k}(\psi )+1}\,\sqrt{N_{q_1}(\psi)}\ \ket{\psi}&\hbox{if }q_j\ne q_k\\
 -N_{q_j}(\psi )\ \ket{\psi}&\hbox{if }q_1=q_k\end{cases}
\end{equation*}
where $N_{q_j}(\psi )$ is the number of $q_j$'s in the basis vector $\psi$. As before, we define the self-adjoint AC-operators
\begin{equation*}
\eta (q_j)=\alpha _ja(q_j)+\alphabar _ja(q_j)^*
\end{equation*}
and the \textit{free boson toy quantum fields}
\begin{equation}         
\label{eq24}
\psi =\sum _{j=1}^m\eta (q_j)=\sum _{j=1}^m\sqbrac{\alpha _ja(q_j)+\alpha _ja(q_j)^*}
\end{equation}

We finally consider a mixed system of fermions $p_1,\ldots ,p_m$ with the same mass and bosons $q_1,\ldots ,q_n$ with the same mass. Again, we limit the total number of particles to $s$. The corresponding Hilbert space $L^{(m,n,s)}$ has orthonormal basis
\begin{equation*}
\ket{p_{j_1}\cdots p_{j_r}q_{k_1}\cdots q_{k_t}}
\end{equation*}
The basis elements are antisymmetric in the $p$'s, symmetric in the $q$'s and symmetric under an interchange of $p$'s and $q$'s. We then have the two free toy quantum fields $\phi$ of \eqref{eq23} and $\psi$ of \eqref{eq24} acting on $L^{(m,n,s)}$.

\section{Eigenstructures for Free Toy Fields} 
One way to understand free toy fields is to find their eigenvalues and eigenvectors which can then be employed to construct their spectral representation. The simplest way to begin is to study the AC-operator
\begin{equation}         
\label{eq31}
\eta (p_1)=\alpha a(p_1)+\alphabar a(p_1)^*
\end{equation}
on the fermion Hilbert space $K^s$. It is easy to check that the eigenvalues of $\eta (p_1)$ are $\pm\ab{\alpha}$. The eigenvectors corresponding to $\ab{\alpha}$ have the form
 \begin{equation}         
\label{eq32}
\ab{\alpha}\ket{p_{i_1}\cdots p_{i_k}}+\alphabar\ket{p_1p_{i_1}\cdots p_{i_k}},\quad i_j\ne 1
\end{equation}
and the eigenvectors corresponding to $-\ab{\alpha}$ have the form
\begin{equation}         
\label{eq33}
\ab{\alpha}\ket{p_{i_1}\cdots p_{i_k}}-\alphabar\ket{p_1p_{i_1}\cdots p_{i_k}},\quad i_j\ne 1
\end{equation}
Since there are $2^{s-1}$ vectors of the form \eqref{eq32} and $2^{s-1}$ vectors of the form \eqref{eq33}, we have a complete set of eigenvectors (which we have not bothered to normalize). Notice that because of the form of the eigenvalues, we have
$\eta (p_1)^2=\ab{\alpha}^2I$.

We next consider the AC-operator
\begin{equation}         
\label{eq34}
\eta (p_2)=\beta a(p_2)+\betabar a(p_2)^*
\end{equation}
whose eigenvalues are $\pm\ab{\beta}$ and whose eigenvectors are similar to \eqref{eq32}, \eqref{eq33}. For concreteness, letting $s=2$, the orthonormal basis for the 4-dimensional Hilbert space $K^2$ is: $\ket{0}$, $\ket{p_1}$, $\ket{p_2}$, $\ket{p_1p_2}$. We consider a free toy field defined by $\phi =\eta (p_1)+\eta (p_2)$. Relative to the given bases we have the matrix representation.
\begin{equation*}
\phi =\begin{bmatrix}\noalign{\smallskip}
0&\alpha&\beta&0\\\noalign{\smallskip}\alphabar&0&0&-\beta\\\noalign{\smallskip}\betabar&0&0&\alpha\\\noalign{\smallskip}
0&-\betabar&\alphabar&0\\\noalign{\smallskip}\end{bmatrix}\end{equation*}
The eigenvalues of $\phi$ are $\pm\sqrt{\ab{\alpha}^2+\ab{\beta}^2}$ and each of these has multiplicity two. The corresponding eigenvectors (which are not orthonormalized) are
\begin{equation*}
\begin{bmatrix}
-\alpha\\\noalign{\smallskip}-\sqrt{\ab{\alpha}^2+\ab{\beta}^2}\\\noalign{\smallskip}0\\\noalign{\smallskip}\betabar\end{bmatrix},\ 
\begin{bmatrix}\noalign{\smallskip}
\sqrt{\ab{\alpha}^2+\ab{\beta}^2}\\\noalign{\smallskip}\alphabar\\\noalign{\smallskip}\betabar\\\noalign{\smallskip}0\end{bmatrix},\ 
\begin{bmatrix}
-\alpha\\\noalign{\smallskip}\sqrt{\ab{\alpha}^2+\ab{\beta}^2}\\\noalign{\smallskip}0\\\noalign{\smallskip}\betabar\end{bmatrix},\ 
\begin{bmatrix}\noalign{\smallskip}
-\sqrt{\ab{\alpha}^2+\ab{\beta}^2}\\\noalign{\smallskip}\alphabar\\\noalign{\smallskip}\betabar\\\noalign{\smallskip}0\end{bmatrix}
\end{equation*}

We next consider the 8-dimensional Hilbert space $K^3$ with orthonormal basis: $\ket{0}$, $\ket{p_1}$, $\ket{p_2}$, $\ket{p_3}$,
$\ket{p_1p_2}$, $\ket{p_1p_3}$, $\ket{p_2p_3}$, $\ket{p_1p_2p_3}$. Relative to this basis we have
\begin{equation*}
\phi=\begin{bmatrix}
0&\alpha&\beta&0&0&0&0&0\\\noalign{\smallskip}\alphabar&0&0&0&-\beta&0&0&0\\\noalign{\smallskip}
\betabar&0&0&0&\alpha&0&0&0\\\noalign{\smallskip}0&0&0&0&0&\alpha&\beta&0\\\noalign{\smallskip}
0&-\betabar&\alphabar&0&0&0&0&0\\\noalign{\smallskip}0&0&0&\alphabar&0&0&0&\beta\\\noalign{\smallskip}
0&0&0&\betabar&0&0&0&\alpha\\\noalign{\smallskip}0&0&0&0&0&\betabar&\alphabar&0\\\noalign{\smallskip}
\end{bmatrix}
\end{equation*}
The eigenvalues of $\phi$ are:
$-\sqrt{\ab{\alpha}^2+\ab{\beta}^2\,}$, $\sqrt{\ab{\alpha}^2+\ab{\beta}^2\,}$, $\ab{\alpha}+\ab{\beta}$,
$-\paren{\ab{\alpha}+\ab{\beta}}$, $\ab{\,\ab{\alpha}-\ab{\beta}\,}$, $-\ab{\,\ab{\alpha}-\ab{\beta}\,}$.
The first two have multiplicity two and last four have multiplicity one. The corresponding (not orthonormalized) eigenvectors are:
\begin{align*}
\begin{bmatrix}
0\\0\\0\\-\alpha\beta\\0\\\ab{\alpha}\beta\\-\alpha\ab{\beta}\\\ab{\alpha}\ab{\beta}\end{bmatrix},\quad
\begin{bmatrix}0\\0\\0\\-\alpha\beta\\0\\-\ab{\alpha}\beta\\\alpha\ab{\beta}\\\ab{\alpha}\ab{\beta}\end{bmatrix},\quad
&\begin{bmatrix}0\\0\\0\\\alpha\beta\\0\\-\ab{\alpha}\beta\\-\alpha\ab{\beta}\\\ab{\alpha}\ab{\beta}\end{bmatrix},\quad
\begin{bmatrix}0\\0\\0\\-\alpha\beta\\0\\\ab{\alpha}\beta\\\alpha\ab{\beta}\\\ab{\alpha}\ab{\beta}\end{bmatrix}\\
\begin{bmatrix}\noalign{\smallskip}
-\sqrt{\ab{\alpha}^2+\ab{\beta}^2\,}\\\alphabar\\\betabar\\0\\0\\0\\0\\0\end{bmatrix},\ 
\begin{bmatrix}-\alpha\\\sqrt{\ab{\alpha}^2+\ab{\beta}^2\,}\\0\\0\\\betabar\\0\\0\\0\end{bmatrix},\ 
&\begin{bmatrix}\noalign{\smallskip}\sqrt{\ab{\alpha}^2+\ab{\beta}^2\,}\\\alphabar\\\betabar\\0\\0\\0\\0\\0\end{bmatrix},\ 
\begin{bmatrix}-\alpha\\-\sqrt{\ab{\alpha}^2+\ab{\beta}^2\,}\\0\\0\\\betabar\\0\\0\\0\end{bmatrix}.\\
\end{align*}

Letting $\eta (p_3)=\gamma a(p_3)+\gammabar a(p_3)^*$ We have the free toy field $\phi =\eta (p_1)+\eta (p_2)+\eta (p_3)$ with matrix representation
\begin{equation*}
\phi=\begin{bmatrix}
0&\alpha&\beta&\gamma&0&0&0&0\\\noalign{\smallskip}\alphabar&0&0&0&-\beta&-\gamma&0&0\\\noalign{\smallskip}
\betabar&0&0&0&\alpha&0&-\gamma&0\\\noalign{\smallskip}\gammabar&0&0&0&0&\alpha&\beta&0\\\noalign{\smallskip}
0&-\betabar&\alphabar&0&0&0&0&\gamma\\\noalign{\smallskip}0&\gammabar&0&\alphabar&0&0&0&\beta\\\noalign{\smallskip}
0&0&-\gammabar&\betabar&0&0&0&\alpha\\\noalign{\smallskip}0&0&0&0&\gammabar&\betabar&\alphabar&0\\\noalign{\smallskip}
\end{bmatrix}
\end{equation*}
The eigenvalues of $\phi$ are:

$\pm\sqrt{\ab{\alpha}^2+\ab{\beta}^2+\ab{\gamma}^2\,}$,
$\pm\sqrt{\ab{\alpha}^2+\ab{\beta}^2+\ab{\gamma}^2\pm 2\sqrt{\ab{\beta}^2(\ab{\alpha}^2+\ab{\gamma}^2\,}\,}$.\newline
The first two have multiplicity two and the others have multiplicity one. The eigenvectors are rather complicated and will be omitted.

We now study free boson toy fields. These are more complicated then the fermion case because of the repeated particles. The simplest nontrivial example is the 6-dimensional symmetric Hilbert space $J^{(2,2)}$ with basis: $\ket{0}$, $\ket{q_1}$, $\ket{q_2}$, $\ket{q_1^2}$,
$\ket{q_1q_2}$, $\ket{q_2^2}$. Letting $\eta (q_1)=\alpha a(q_1)+\alphabar a(q_1)^*$ and $\eta (q_2)=\beta a(q_2)+\betabar a(q_2)^*$ we have the matrix representation of the free toy boson field $\phi =\eta (q_1)+\eta (q_2)$ given by
\begin{equation*}
\phi=\begin{bmatrix}
0&\alpha&\beta&0&0&0\\\noalign{\smallskip}\alphabar&0&0&\sqrt{2}\alpha&\beta&0\\\noalign{\smallskip}
\betabar&0&0&0&\alpha&\sqrt{2}\beta\\\noalign{\smallskip}0&\sqrt{2}\,\alphabar&0&0&0&0\\\noalign{\smallskip}
0&\betabar&\alphabar&0&0&0\\\noalign{\smallskip}0&0&\sqrt{2}\,\betabar&0&0&0\\\noalign{\smallskip}
\end{bmatrix}
\end{equation*}
The eigenvalues are: $0$, $\pm\sqrt{\ab{\alpha}^2+\ab{\beta}^2\,}$, $\pm\sqrt{3}\sqrt{\ab{\alpha}^2+\ab{\beta}^2\,}$. The eigenvalue $0$ has multiplicity two and the others have multiplicity one. The corresponding eigenvectors are:
\begin{align*}
\begin{bmatrix}\noalign{\smallskip}
-\sqrt{2\,}\alpha\beta\\0\\0\\\alphabar\beta\\0\\\alpha\betabar\end{bmatrix},\quad
&\begin{bmatrix}\noalign{\smallskip}-\sqrt{2\,}\alpha ^2\\0\\0\\\ab{\alpha}^2-\ab{\beta}^2\\\noalign{\smallskip}
\sqrt{2\,}\alpha\betabar\\0\end{bmatrix},\quad
\begin{bmatrix}\noalign{\smallskip}0\\-\beta\sqrt{\ab{\alpha}^2+\ab{\beta}^2\,}\\\noalign{\smallskip}
\alpha\sqrt{\ab{\alpha}^2+\ab{\beta}^2\,}\\\noalign{\smallskip}-\sqrt{2}\,\alphabar\beta\\
\noalign{\smallskip}\ab{\alpha}^2-\ab{\beta}^2\\\noalign{\smallskip}\sqrt{2}\,\alpha\betabar\end{bmatrix},\\
\begin{bmatrix}\noalign{\smallskip}
0\\\beta\sqrt{\ab{\alpha}^2+\ab{\beta}^2\,}\\\noalign{\smallskip}-\alpha\sqrt{\ab{\alpha}^2+\ab{\beta}^2\,}\\\noalign{\smallskip}
-\sqrt{2}\,\alphabar\beta\\\noalign{\smallskip}\ab{\alpha}^2-\ab{\beta}^2\\\noalign{\smallskip}\sqrt{2}\,\alpha\betabar\end{bmatrix},\quad
&\begin{bmatrix}\noalign{\smallskip}\ab{\alpha}^2+\ab{\beta}^2\\\noalign{\smallskip}\sqrt{3}\,\alphabar\sqrt{\ab{\alpha}^2+\ab{\beta}^2\,}\\
\noalign{\smallskip}\sqrt{3}\,\betabar\sqrt{\ab{\alpha}^2+\ab{\beta}^2\,}\\\noalign{\smallskip}\sqrt{2}\,(\alphabar )^2\\
\noalign{\smallskip}2\alphabar\betabar\\\noalign{\smallskip}\sqrt{2}\,(\betabar )^2\end{bmatrix},\quad
\begin{bmatrix}\noalign{\smallskip}\ab{\alpha}^2+\ab{\beta}^2\\\noalign{\smallskip}-\sqrt{3}\,\alphabar\sqrt{\ab{\alpha}^2+\ab{\beta}^2\,}\\
\noalign{\smallskip}-\sqrt{3}\,\betabar\sqrt{\ab{\alpha}^2+\ab{\beta}^2\,}\\\noalign{\smallskip}\sqrt{2}\,(\alphabar )^2\\
\noalign{\smallskip}2\alphabar\betabar\\\noalign{\smallskip}\sqrt{2}\,(\betabar )^2\\
\end{bmatrix}.
\end{align*}

We next consider that 10-dimension symmetric Hilbert space $J^{(2,3)}$ with basis: $\ket{0}$, $\ket{q_1}$, $\ket{q_2}$, $\ket{q_1^2}$,
$\ket{q_1q_2}$, $\ket{q_2^2}$, $\ket{q_1^3}$, $\ket{q_1^2q_2}$, $\ket{q_1q_2^2}$, $\ket{q_2^3}$. The toy boson field
$\phi =\eta (q_1)+\eta (q_2)$ has matrix representation
\begin{equation*}
\phi=\begin{bmatrix}
0&\alpha&\beta&0&0&0&0&0&0&0\\\noalign{\smallskip}\alphabar&0&0&\sqrt{2}\,\alpha&\beta&0&0&0&0&0\\\noalign{\smallskip}
\betabar&0&0&0&\alpha&\sqrt{2}\,\beta&0&0&0&0\\\noalign{\smallskip}0&\sqrt{2}\,\alphabar&0&0&0&0&\sqrt{3}\,\alpha&\beta&0&0\\
\noalign{\smallskip}
0&\betabar&\alphabar&0&0&0&0&\sqrt{2}\,\alpha&\sqrt{2}\,\beta&0\\\noalign{\smallskip}
0&0&\sqrt{2}\,\betabar&0&0&0&0&0&\alpha&\sqrt{3}\,\beta\\\noalign{\smallskip}
0&0&0&\sqrt{3}\,\alphabar&0&0&0&0&0&0\\\noalign{\smallskip}
0&0&0&\betabar&\sqrt{2}\,\alphabar&0&0&0&0&0\\\noalign{\smallskip}
0&0&0&0&\sqrt{2}\,\betabar&\alphabar&0&0&0&0\\\noalign{\smallskip}
0&0&0&0&0&\sqrt{3}\,\betabar&0&0&0&0\\\noalign{\smallskip}
\end{bmatrix}
\end{equation*}
The eigenvalues are:

$0$, $\pm\sqrt{\ab{\alpha}^2+\ab{\beta}^2\,}$, $\pm\sqrt{3}\,\sqrt{\ab{\alpha}^2+\ab{\beta}^2\,}$,
$\pm\sqrt{3\pm\sqrt{6\,}\,}\sqrt{\ab{\alpha}^2+\ab{\beta}^2\,}$.\newline
The eigenvalue $0$ has multiplicity two. We omit the rather complicated eigenvectors.

The most interesting and important of the free toy fields are the mixed ones. The simplest nontrivial example is when we have two fermions $p_1$, $p_2$, two bosons $q_1$, $q_2$ and we let $s=2$. The mixed Hilbert space $L^{(2,2,2)}$ has basis: $\ket{0}$, $\ket{p_1}$,
$\ket{p_2}$, $\ket{q_1}$, $\ket{q_2}$, $\ket{p_1p_2}$, $\ket{p_1q_1}$, $\ket{p_1q_2}$, $\ket{p_2q_1}$, $\ket{p_2q_2}$, $\ket{q_1^2}$,
$\ket{q_1q_2}$, $\ket{q_2^2}$. Letting $\eta (p_1)$, $\eta (p_2)$ have the form \eqref{eq31}, \eqref{eq34}, we obtain the toy fermion field
$\phi =\eta (p_1)+\eta (p_2)$. The eigenvalues are: $0$ (multiplicity 5), $\pm\sqrt{\ab{\alpha}^2+\ab{\beta}^2\,}$ (multiplicity 4, each). In a similar way we define $\eta (q_1)=\gamma a(q_1)+\gammabar a(q_1)^*$ and $\eta (q_2)=\delta a(q_2)+\deltabar a(q_2)^*$. We then obtain the free toy boson field $\psi =\eta (q_1)+\eta (q_2)$. The eigenvalues for $\psi$ are: $0$ (multiplicity 5),
$\pm\sqrt{\ab{\gamma}^2+\ab{\delta}^*\,}$ (multiplicity 3, each) and $\pm\sqrt{3}\sqrt{\ab{\gamma}^2+\ab{\delta}^2\,}$ (multiplicity 1, each). To save space, we omit the matrix representations and eigenvectors.

\section{Interacting Toy Fields} 
Let $\phi =\eta (p_1)+\cdots +\eta (p_n)$ and $\psi =\eta (q_1)+\cdots +\eta (q_m)$ be free toy quantum fields acting on the same Hilbert space $K$. The operators $\phi$ and $\psi$ can be either fermion or boson operators. These fields can interact in various ways, but we shall only consider the simplest nontrivial interaction $\phi\psi$. We shall call the self-adjoint operator 
\begin{equation}         
\label{eq41}
\tau =\tfrac{1}{2}(\phi\psi +\psi\phi )=\tfrac{1}{2}\brac{\phi ,\psi}
\end{equation}
an \textit{interaction toy quantum field}. Even in small dimensional Hilbert spaces, the eigenstructure of $\tau$ can be quite complicated and we shall only consider some simple examples. We have seen that for a fermion AC-operator $\eta (p_1)$ the self-interaction 
$\eta (p_1)^2=\ab{\alpha}^2I$ and we see that for $\phi =\eta (p_1)+\eta (p_2)$ on $K^2$ that $\phi ^2=\paren{\ab{\alpha}^2+\ab{\beta}^2}I$ so this self-interaction is also trivial. However, if $\phi =\eta (p_1)+\eta (p_2)$ on $K^3$ that we previously considered the self-interaction
$\phi ^2$ has eigenvalues $\ab{\alpha}^2+\ab{\beta}^2$ (multiplicity 4), $\paren{\ab{\alpha}-\ab{\beta}}^2$ (multiplicity 2) and 
$\paren{\ab{\alpha}+\ab{\beta}}^2$ (multiplicity 2).

An important case of an interaction toy field comes from the last example of Section~3. Corresponding to free fields $\phi$ and $\psi$ we obtain the interaction toy field \eqref{eq41} whose matrix representation is
\medskip

$\tau=\left [
\begin{array}{*{20}c}
0&0&0&0&0&0&\alpha\gamma&\alpha\delta&\beta\gamma&\beta\delta&0&0&0\\
0&0&0&\alphabar\gamma&\alphabar\delta&0&0&0&0&0&0&0&0\\
0&0&0&\betabar\gamma&\betabar\delta&0&0&0&0&0&0&0&0\\
0&\alpha\gammabar&\beta\gammabar&0&0&0&0&0&0&0&0&0&0\\
0&\alpha\deltabar&\beta\deltabar&0&0&0&0&0&0&0&0&0&0\\\noalign{\smallskip}
0&0&0&0&0&0&\tfrac{-\betabar\gamma}{2}&\tfrac{-\betabar\delta}{2}&\tfrac{\alphabar\,\gamma}{2}
   &\tfrac{\alphabar\,\delta}{2}&0&0&0\\\noalign{\smallskip}
\alphabar\,\gammabar&0&0&0&0&\tfrac{-\beta\gammabar}{2}&0&0&0&0&\tfrac{\alphabar\,\gamma}{\sqrt{2}}&\tfrac{\alphabar\,\delta}{2}
   &0\\\noalign{\smallskip}
\alphabar\,\deltabar&0&0&0&0&\tfrac{-\beta\deltabar}{2}&0&0&0&0&0&\tfrac{\alphabar\,\gamma}{2}&\tfrac{\alphabar\,\delta}{\sqrt{2}}
   \\\noalign{\smallskip}
\betabar\,\gammabar&0&0&0&0&\tfrac{\alpha\gammabar}{2}&0&0&0&0&\tfrac{\betabar\,\gamma}{\sqrt{2}}&\tfrac{\betabar\,\delta}{2}
   &0\\\noalign{\smallskip}
\betabar\,\deltabar&0&0&0&0&\tfrac{\alpha\deltabar}{2}&0&0&0&0&0&\tfrac{\betabar\,\gamma}{2}&\tfrac{\betabar\,\delta}{\sqrt{2}}
   \\\noalign{\smallskip}
0&0&0&0&0&0&\tfrac{\alpha\gammabar}{\sqrt{2}}&0&\tfrac{\beta\gammabar}{\sqrt{2}}&0&0&0&0\\\noalign{\smallskip}
0&0&0&0&0&0&\tfrac{\alpha\deltabar}{2}&\tfrac{\alpha\gammabar}{2}&\tfrac{\beta\deltabar}{2}&\tfrac{\beta\gammabar}{2}&0
  &0&0\\\noalign{\smallskip}
0&0&0&0&0&0&0&\tfrac{\alpha\deltabar}{\sqrt{2}}&0&\tfrac{\beta\deltabar}{\sqrt{2}}&0&0&0\\\noalign{\smallskip}
\end{array}
\right ]$\medskip

\noindent The eigenvalues are: $0$ (multiplicity 5),
$\pm\sqrt{\ab{\alpha}^2+\ab{\beta}^2\,}\sqrt{\ab{\gamma}^2+\ab{\delta}^2\,}$ (multiplicity 1, each),
$\pm\tfrac{1}{2}\sqrt{\ab{\alpha}^2+\ab{\beta}^2\,}\sqrt{\ab{\gamma}^2+\ab{\delta}^2\,}$ (multiplicity 2, each)
and $\pm\sqrt{\tfrac{3}{2}}\,\sqrt{\ab{\alpha}^2+\ab{\beta}^2\,}\sqrt{\ab{\gamma}^2+\ab{\delta}^2\,}$ (multiplicity 1, each). We omit the corresponding eigenvectors.

We now begin an analysis of interaction toy fields. We admit that this is preliminary work and much more needs to be done. Consider the very simple field
\begin{equation}        
\label{eq42}
\phi=\tfrac{1}{2}\sqbrac{\eta (p_1)\eta (q_1)+\eta (q_1)\eta (p_1)}
\end{equation}
We assume that $\eta (p_1)$ and $\eta (q_1)$ are boson AC-operators acting on the boson Hilbert space $J^{(n,m,s)}$ where the particles $p_1\ldots ,p_n$ are bosons of mass $m_1$ and $q_1,\ldots ,q_m$ are bosons of mass $m_2$. Moreover, we assume that $\eta (p_1)$ and
$\eta (q_1)$ have the form:
\begin{align*}
\eta (p_1)&=\alpha a(p_1)+\alphabar a(p_1)^*\\
\eta (q_1)&=\beta a(q_1)+\betabar a(q_1)^*
\end{align*}
The general basis element for $J^{(n,m,s)}$ is:
\begin{equation}        
\label{eq43}
\ket{p_1^{r_1}\cdots p_n^{r_n}q_1^{s_1}\cdots q_m^{s_m}}
\end{equation}
where $\sum r_i+\sum s_i\le s$. Because of the structure of $\phi$ in \eqref{eq42}, we are mainly concerned with $p_1,q_1$ and we use the notation $\ket{rs}$ for a vector of the form \eqref{eq43}. When we write $\ket{r_1s_1}\sim\ket{r_2s_2}$, we mean that $r_1\ne r_2$ or $s_1\ne s_2$ but that the other vectors in \eqref{eq43} are identical. For example,
\begin{equation*}
\ket{p_1p_2q_1^2q_2}\sim\ket{p_1p_2q_1q_2}
\end{equation*}
If a vector has a representation
\begin{equation*}
\psi =\sum _{k=1}^t\beta _k\ket{i_kj_k},\quad\ket{i_rj_r}\sim\ket{i_sj_s},\quad r\ne s,\quad \beta _k\in\complex
\end{equation*}
we say that $\psi$ has \textit{type} $t$ and \textit{form} $\paren{\ket{i_1j_1},\ldots ,\ket{i_tj_t}}$. If $i_k+j_k$ is even for all $k$, then $\psi$ has \textit{even form} and if $i_k+j_k$ is odd for all $k$, then $\psi$ has \textit{odd form}. For example, $\ket{00}$ has even form of type~1,
$\beta _1\ket{01}+\beta _2\ket{10}$ has odd form of type~2 and 
\begin{equation*}
\beta _1\ket{00}+\beta _2\ket{11}+\beta _3\ket{20}+\beta _4\ket{02}
\end{equation*}
has even form of type~4. We shall employ this terminology to classify the eigenvectors of $\phi$. It is useful to observe that
\begin{align}        
\label{eq44}
\phi&=\alpha\beta a(p_1)a(q_1)+\alphabar\,\betabar a(p_1)^*a(q_1)^*+\tfrac{1}{2}\,\alpha\betabar\brac{a(p_1),a(q_1)^*}\notag\\
   &\qquad +\tfrac{1}{2}\,\alphabar\,\beta\brac{a(p_1)^*,a(q_1)}
\end{align}
The simplest case is the boson Hilbert space $J^{(1,1,2)}$ with $s=2$ and bosons $p_1,q_1$. The basis for $J^{(1,1,2)}$ is:
$\ket{0}$, $\ket{p_1}$, $\ket{q_1}$, $\ket{p_1q_1}$, $\ket{p_1^2}$, $\ket{q_1^2}$. Applying \eqref{eq44} the matrix representation for $\phi$ becomes:
\begin{equation*}
\phi=\begin{bmatrix}
0&0&0&\alpha\beta&0&0\\\noalign{\smallskip}0&0&\alphabar\,\beta&0&0&0\\\noalign{\smallskip}
0&\alpha\,\betabar&0&0&0&0\\\noalign{\smallskip}\alphabar\,\betabar&0&0&0&\tfrac{\alpha\betabar}{\sqrt{2}}
  &\tfrac{\alphabar\,\beta}{\sqrt{2}}\\\noalign{\smallskip}
0&0&0&\tfrac{\alphabar\,\beta}{\sqrt{2}}&0&0\\\noalign{\smallskip}0&0&0&\tfrac{\alpha\betabar}{\sqrt{2}}&0&0\\\noalign{\smallskip}
\end{bmatrix}
\end{equation*}
The eigenvalues of $\phi$ are: $0$ (multiplicity 2), $\pm\ab{\alpha}\ab{\beta}$, $\pm\sqrt{2}\,\ab{\alpha}\ab{\beta}$. The corresponding eigenvectors are:
\begin{equation*}
\begin{bmatrix}
-\beta\\0\\0\\0\\0\\\sqrt{2\,}\,\betabar\end{bmatrix},
\begin{bmatrix}-\alpha\\0\\0\\0\\\sqrt{2\,}\alphabar\\0\end{bmatrix},
\begin{bmatrix}0\\\ab{\alpha}\beta\\\noalign{\smallskip}\alpha\ab{\beta}\\\noalign{\smallskip}0\\0\\0\end{bmatrix},
\begin{bmatrix}0\\-\ab{\alpha}\beta\\\noalign{\smallskip}\alpha\ab{\beta}\\\noalign{\smallskip}0\\0\\0\end{bmatrix},
\begin{bmatrix}\sqrt{2}\,\beta/\,\betabar\\0\\0\\
2\beta\ab{\alpha}/\alpha\ab{\beta}\\\noalign{\smallskip}\alphabar\,\beta/\alpha\betabar\\\noalign{\smallskip}1\end{bmatrix},
\begin{bmatrix}\sqrt{2}\,\beta/\,\betabar\\0\\0\\
-2\beta\ab{\alpha}/\alpha\ab{\beta}\\\noalign{\smallskip}\alphabar\,\beta/\alpha\betabar\\\noalign{\smallskip}1\end{bmatrix}.
\end{equation*}
We see that the eigenvectors have types~2,~4 and form
\begin{align*}
\paren{\ket{00},\ket{02}},\ \paren{\ket{00},\ket{20}},&\ \paren{\ket{10},\ket{01}},\ \paren{\ket{10},\ket{01}},\\
\paren{\ket{00},\ket{11},\ket{20},\ket{02}},&\ \paren{\ket{00},\ket{11},\ket{20},\ket{02}}.
\end{align*}
The third and fourth have odd form while the others have even form.

A slightly more involved case is the boson Hilbert space $J^{(2,1,2)}$ with bosons $p_1,p_2,q_1$ and $s=2$. The basis for $J^{(2,1,2)}$ is:
$\ket{0}$, $\ket{p_1}$, $\ket{p_2}$, $\ket{q_1}$, $\ket{p_1^2}$, $\ket{p_2^2}$, $\ket{p_1p_2}$, $\ket{p_1q_1}$, $\ket{p_2q_1}$, $\ket{q_1^2}$.
We omit the matrix representation of $\phi$ but we list its eigenvalues: $0$ (multiplicity 4), $\pm\ab{\alpha}\ab{\beta}$,
$\pm\tfrac{1}{2}\,\ab{\alpha}\ab{\beta}$, $\pm\sqrt{2\,}\,\ab{\alpha}\ab{\beta}$. The corresponding eigenvectors are:
\begin{align*}
\begin{bmatrix}
-\beta\\0\\0\\0\\0\\0\\0\\0\\0\\\sqrt{2\,}\,\betabar\end{bmatrix},\quad
\begin{bmatrix}-\alpha\\0\\0\\0\\\sqrt{2\,}\alphabar\\0\\0\\0\\0\\0\end{bmatrix},\quad
&\begin{bmatrix}0\\0\\0\\0\\0\\1\\0\\0\\0\\0\end{bmatrix},\ 
\begin{bmatrix}0\ \\0\\1\\0\\0\\0\\0\\0\\0\\0\end{bmatrix},\ 
\begin{bmatrix}0\\\alphabar\,\beta\\0\\\ab{\alpha}\ab{\beta}\\0\\0\\0\\0\\0\\0\end{bmatrix},\quad
\begin{bmatrix}0\\-\alphabar\,\beta\\0\\\ab{\alpha}\ab{\beta}\\0\\0\\0\\0\\0\\0\end{bmatrix},\\
\begin{bmatrix}0\\0\\0\\0\\0\\0\\\alphabar\,\beta\\0\\\ab{\alpha}\ab{\beta}\\0\end{bmatrix},\quad
\begin{bmatrix}0\\0\\0\\0\\0\\0\\-\alphabar\,\beta\\0\\\ab{\alpha}\ab{\beta}\\0\end{bmatrix},\quad
&\begin{bmatrix}\noalign{\smallskip}\sqrt{2}\,\beta/\,\betabar\\0\\0\\0\\
\alphabar\,\beta/\alpha\betabar\\0\\0\\2\alphabar\,\beta/\ab{\alpha}\ab{\beta}\\0\\1\end{bmatrix},\quad
\begin{bmatrix}\noalign{\smallskip}\sqrt{2}\,\beta/\,\betabar\\0\\0\\0\\
\alphabar\,\beta/\alpha\betabar\\0\\0\\-2\alphabar\,\beta/\ab{\alpha}\ab{\beta}\\0\\1\end{bmatrix}.
\end{align*}
The eigenvectors have types~1,~2,~4, and even form $\paren{\ket{00}}$, $\paren{\ket{00},\ket{02}}$, $\paren{\ket{00},\ket{20}}$,
$\paren{\ket{00},\ket{11},\ket{20},\ket{02}}$, and odd form $\paren{\ket{10},\ket{01}}$.

The fermion case is much simpler than the boson case. Our last example is a fermion Hilbert space $K^{(2,2,4)}$ with two fermions $p_1,p_2$ of mass $m_1$, two other fermions $q_1,q_2$ of mass $m_2$ and $s=4$. This 16-dimensional space has basis:
$\ket{0}$, $\ket{p_1}$, $\ket{p_2}$, $\ket{q_1}$, $\ket{q_2}$, $\ket{p_1p_2}$, $\ket{p_1q_1}$, $\ket{p_1q_2}$, $\ket{p_2q_1}$,
$\ket{p_2q_2}$, $\ket{q_1q_2}$, $\ket{p_1p_2q_1}$, $\ket{p_1p_2q_2}$, $\ket{p_1q_1q_2}$, $\ket{p_2q_1q_2}$, $\ket{p_1p_2q_1q_2}$.
We define the AC-operators:
\begin{align*}
\eta (p_1)&=\alpha a(p_1)+\alphabar a(p_1)^*\\
\eta (p_2)&=\beta a(p_2)+\betabar a(p_2)^*\\
\eta (q_1)&=\gamma a(q_1)+\gammabar a(q_1)^*\\
\eta (q_2)&=\delta a(q_2)+\deltabar a(q_2)^*\\
\end{align*}
The corresponding free toy fermion fields become:
\begin{align*}
\phi&=\eta (p_1)+\eta (p_2)\\
\psi&=\eta (q_1)+\eta (q_2)\\
\end{align*}
The interaction toy field is $\tau =\tfrac{1}{2} (\phi\psi +\psi\phi )$. we omit the matrix representation of $\tau$ except for saying that the first row of $\tau$ is
\begin{equation*}
\sqbrac{0\cdots 0\ \alpha\gamma\ \alpha\delta\ \beta\gamma\ \beta\delta\ 0\cdots 0}
\end{equation*}
and the other rows are permutations of the first row with various minus signs and complex conjugations. The eigenvalues of $\tau$ are surprisingly simple. Letting $\omega _1=\sqrt{\ab{\alpha}^2+\ab{\beta}^2\,}$ and $\omega _2=\sqrt{\ab{\gamma}^2+\ab{\delta}^2\,}$, the eigenvalues are $\pm\omega _1\omega _2$ each with multiplicity 8. One of the eigenvectors has the form
\begin{equation*}
\sqbrac{0\cdots 0\ -\gamma\omega _1\ -\beta\omega _2\ 0\ \alpha\omega _2\ 0\cdots 0\ \deltabar\,\omega _1}^T
\end{equation*}
and the others are permutations of this one with various minus signs and complex conjugations.

This pattern continues. For example, for fermions $p_1,p_2,p_3$ and $q_1,q_2,q_3$ we obtain the 64-dimensional Hilbert space
$K^{(3,3,6)}$. With the obvious notation, the eigenvalues of $\tau$ become $\pm\omega _1\omega _2\omega _3$ each with multiplicity 32.

\section{Discrete Quantum Field Theory} 
This section presents the background for a discrete quantum field theory \cite{bdp16,cro16,gud172,hoo14}. Toy models emerge from this theory by imposing a particle number cut-off. We then see how the work of previous sections can be applied to approximate this general discrete theory. As before, we shall neglect spin so we are actually considering scalar fields. The main goal of this section is to introduce the concept of toy scattering operators.

Our basic assumption is that spacetime is discrete and has the form of a 4-dimensional cubic lattice $\sscript$ \cite{gud16,gud171,gud172}. We then have $\sscript =\integers ^+\times\integers ^3$ where $\integers ^+=\brac{0,1,2,\ldots}$ represents discrete time and $\integers =\brac{0,\pm 1,\pm 2,\ldots}$ so $\integers ^3$ represents discrete 3-space. If $x=(x_0,x_1,x_2,x_3)\in\sscript$, we sometimes write $x=(x_0,\bfx )$ where $x_0\in\integers ^+$ is time and $\bfx\in\integers ^3$ is a 3-space point. We equip $\sscript$ with the Minkowski distance
\begin{equation*}
\doubleab{x}_4^2=x_0^2-\doubleab{\bfx}_3^2=x_0^2-x_1^2-x_2^2-x_3^2
\end{equation*}
where we use units in which the speed of light is 1.

The dual of $\sscript$ is denoted by $\sscripthat$. We regard $\sscripthat$ as having the identical structure as $\sscript$ except we denote elements of $\sscripthat$ by
\begin{equation*}
p=(p_0,\bfp )=(p_0,p_1,p_2,p_3)
\end{equation*}
and interpret $p$ as the energy-momentum vector for a particle. In fact, we sometimes call $p\in\sscripthat$ a particle. Moreover, we only consider the forward cone
\begin{equation*}
\brac{p\in\sscripthat\colon\doubleab{p}_4\ge 0}
\end{equation*}
For a particular $p\in\sscripthat$ we call $p_0\ge 0$ the \textit{total energy}, $\doubleab{\bfp}_3\ge 0$ the \textit{kinetic energy} and $m=\doubleab{p}_4$ the \textit{mass} of $p$. The integers $p_1,p_2,p_3$ are \textit{momentum components}. Since
\begin{equation*}
m^2=\doubleab{p}_4^2=p_0^2-\doubleab{\bfp}_3^2
\end{equation*}
we conclude that Einstein's energy formula $p_0=\sqrt{m^2+\doubleab{\bfp}_3^2\,}$ holds. For mass $m$, we define the \textit{mass hyperboloid} by
\begin{equation*}
\Gamma _m=\brac{p\in\sscripthat\colon\doubleab{p}_4=m}
\end{equation*}
Moreover, for $x\in\sscript$, $p\in\sscripthat$ we define the indefinite inner product
\begin{equation*}
px=p_0x_0-p_1x_1-p_2x_2-p_3x_3
\end{equation*}
We now define three Hilbert spaces. For fermions $p_1,p_2,\ldots\,$, we define the \textit{fermion Hilbert space} $K$ to be the separable antisymmetric Hilbert space with orthonormal basis
\begin{equation*}
\ket{0},\ket{p_1},\ket{p_2},\ldots ,\ket{p_1p_2},\ket{p_1p_3},\cdots ,\ket{p_1p_2\cdots p_n},\cdots
\end{equation*}
For bosons $q_1,q_2,\ldots$ we define the \textit{boson Hilbert space} $J$ to be the symmetric Hilbert space with orthonormal basis
\begin{equation*}
\ket{0},\ket{q_1},\ket{q_2},\ldots ,\ket{q_1^2},\ket{p_1p_2},\cdots ,\ket{q_1^{r_1}q_2^{r_2}\cdots q_n^{r_n}},\cdots
\end{equation*}
For fermions $p_1,p_2,\ldots$ and bosons $q_1,q_2,\ldots\,$, we define the \textit{mixed Hilbert space} $L$ analogous to the construction in Section~3. Moreover, we define the annihilation and creation operators $a(p_j)$, $a(q_j)$, $a(p_j)^*$, $a(q_j)^*$ as we did in Section~2. On the Hilbert space $K$, for any $x\in\sscript$, $r\in\integers ^+$, we define the \textit{free fermion quantum field} at $x$ with mass $m$ and maximal total energy $r$ \cite{gud16,gud171,gud172} by:
\begin{equation}        
\label{eq51}
\phi (x,r)=\sum\brac{\tfrac{1}{p_0}\sqbrac{a(p)e^{i\pi px/2}+a(p)^*e^{-i\pi px/2}}\colon p\in\Gamma _m,p_0\le r}
\end{equation}
Since there are only a finite number of terms in the summation in \eqref{eq51}, we see that $\phi (x,r)$ is a bounded self-adjoint operator on
$K$. We define free boson quantum fields on $J$ in an analogous way. We can also define quantum fields for particles of various masses.

We now impose the number of particles cut-off $s$ to obtain the finite-dimensional subspaces $K^s$, $J^s$, $L^s$ of $K$, $J$, $L$ considered in Section~2. The restriction of free field $\phi (x,r)$ to one of these subspaces becomes a free toy field and is denoted by
$\phi ^s(x,r)$. Letting $s\to\infty$, we have in a certain weak sense that
\begin{equation*}
\lim _{s\to\infty}\phi ^s(x,r)=\phi (x,r)
\end{equation*}
As shown in \cite{gud171,gud172} we can also relax the maximal energy restriction by letting $\phi (x)=\lim\limits _{r\to\infty}\phi (x,r)$.

For concreteness, consider the fermion Hilbert space $K$ and the other two cases are similar. Let $H(x_0)$ be a self-adjoint operator on $K$ that is a function of time $x_0\in\integers ^+$. This operator is specified by the theory and we call it a \textit{Hamiltonian}. The operator $H(x_0)$ is usually derived from a \textit{Hamiltonian density} $\hscript (x)$, $x\in\sscript$, where $\hscript (x)$ are again self-adjoint operators on $K$. The \textit{space-volume} at $x_0$ is given by the cardinality $V(x_0)=\ab{\brac{x\colon\doubleab{\bfx}_3\le x_0}}$ and we define
\begin{equation}        
\label{eq52}
H(x_0)=\frac{1}{V(x_0)}\sum\brac{\hscript (x)\colon\doubleab{\bfx}_3\le x_0}
\end{equation}
For a Hamiltonian $H(x_0)$ the corresponding \textit{scattering operator} at time $x_0$ is the unitary operator $S(x_0)=e^{iH(x_0)}$. The \textit{final scattering operator} is defined as $S=\lim\limits _{x_0\to\infty}S(x_0)$ when this limit exists. The scattering operator is used to find scattering amplitudes and probabilities. For example, suppose that two particles with energy-momentum $p$ and $q$ collide at time $0$. We would like to find the probability that their energy-momentum is $p'$ and $q'$ at time $x_0$. (By conservation of energy-momentum, we usually assume that $p'+q'=p+q$.) The axioms of quantum mechanics tell us that the scattering amplitude for this interaction is
$\bra{p'q'}S(x_0)\ket{pq}$ and the probability becomes
\begin{equation*}
\ab{\bra{p'q'}S(x_0)\ket{pq}}^2
\end{equation*}

A fundamental question in quantum field theory is: ``How do we find the Hamiltonian density $\hscript (x)$, $x\in\sscript$?'' The answer is usually that $\hscript (x)$ is determined by the interaction of quantum fields. But the problem now is that we don't know how the fields interact and we frequently have to make an intelligent guess. Suppose $\phi (x,r)$ and $\psi (x,r)$ are quantum fields. As mentioned in Section~4, probably the simplest nontrivial interaction is $\phi (x,r)\psi (x,r)$ so we form the self-adjoint operator
\begin{equation*}
\tau (x,r)=\tfrac{1}{2}\brac{\phi (x,r),\psi (x,r)}
\end{equation*}
To simplify the calculations we can consider the toy interaction field
\begin{equation*}
\tau ^s(x,r)=\tfrac{1}{2}\brac{\phi ^s(x,r),\psi ^s(x,r)}
\end{equation*}
discussed in Section~4. We then define the \textit{toy Hamiltonian density} to be $\tau ^s(x,r)$. This quantity has the extra parameter $r$ which we can let approach infinity at the end of the calculation. We now construct the toy Hamiltonian
\begin{equation*}
H^s(x_0,r)=\tfrac{1}{V(x_0)}\sum\brac{\tau ^s(x,r)\colon\doubleab{\bfx}_3\le x_0}
\end{equation*}
The resulting toy scattering operator becomes
\begin{equation}        
\label{eq53}
S^s(x_0,r)=e^{iH^s(x_0,r)}
\end{equation}

In Sections~3 and 4 we stressed the importance of finding the eigenstructure of toy free and interaction quantum fields. Once we have calculated the eigenstructure of $H^s(x_0,r)$ we can write its spectral representation
\begin{equation}        
\label{eq54}
H^s(x_0,r)=\sum _{j=1}^n\lambda _jP_j
\end{equation}
where $\lambda _j$ are the distinct eigenvalues of $H^s(x_0,r)$ and $P_j$ is the orthogonal projection onto the eigenspace of $\lambda _j$, $j=1,\ldots ,n$. In this case, $P_jP_k=\delta _{jk}P_j$ and $\sum\limits _{j=1}^nP_j=I$. Applying \eqref{eq53}, \eqref{eq54}, we obtain a complete formula for $S^s(x_0,r)$ given by
\begin{equation*}
S^s(x_0,r)=\sum _{j=1}^ne^{i\lambda _j}P_j
\end{equation*}
Of course, the examples of eigenstructures computed in Sections~3 and 4 were much simpler than that of $H^s(x_0,r)$ and computer assistance must be employed for the latter. It is clear that the present work is only the beginning and much more must be done to develop a compete theory. Once this is accomplished, we can check the theory against experimental results.


\begin{thebibliography}{99}
\bibitem{bdp16}A.~Bisco, G.~D'Ariano and P.~Perinotti, Special relativity in a discrete quantum universe,
arXiv: quant-ph 1503.01017v3 (2016).
\bibitem{cro16}D.~Crouse, On the nature of discrete space-time, arXiv: quant-ph 1608.08506v1 (2016).
\bibitem{gud16}S.~Gudder, Discrete scalar quantum field theory, arXiv: gen-ph 1610.07877v1 (2016).
\bibitem{gud171}S.~Gudder, Reconditioning in discrete quantum field theory, \textit{Int.\ J.\ Theor.\ Phys.}
DOI 10.1007/s10773-017-3350-6 (2017).
\bibitem{gud172}S.~Gudder, Discrete spacetime quantum field theory, arXiv: gen-ph 1704.01639 (2017).
\bibitem{hag14}A.~Hagar, \textit{Discrete or continuous?: the quest for fundamental length in modern physics}, Cambridge University Press (2014).
\bibitem{hoo14}G.~'t Hooft, Relating the quantum mechanics  of discrete systems to standard canonical quantum mechanics, \textit{Found. Phys.} \textbf{44}, 406--425 (2014).
\end{thebibliography}
\end{document}